# Dielectric and structural studies of ferroelectric phase evolution in dipole pair substituted barium titanate ceramics

Vignaswaran K.Veerapandiyan [a, †, *]; Marco Deluca [b]; Scott T. Misture [a]; Walter A. Schulze [a]; Steven M. Pilgrim [a]; Steven C. Tidrow [a, *]

[a] New York State College of Ceramics, Alfred University, Alfred 14802, New York, U.S.A.

[b] Materials Center Leoben Forschung GmbH, Roseggerstrasse 12, Leoben 8700, Austria.

† Present address: Materials Center Leoben Forschung GmbH, Roseggerstrasse 12, Leoben 8700, Austria.

* Authors to whom correspondence should be addressed. E-mail: tidrow@alfred.edu and vignaswaran.veerapandiyan@mcl.at.

**Abstract:**

Ba{[Ga$_x$,Ta$_x$]Ti$_{(1-2x)}$}O$_3$ ceramics with x equal to 0, 0.0025, 0.005, 0.01, 0.025 and 0.05 have been prepared by conventional solid-state reaction. Structural and dielectric characterization have been performed to investigate the effect of dipole-pair substitution concentration on the macroscopic dielectric properties. Ba{[Ga$_x$,Ta$_x$]Ti$_{(1-2x)}$}O$_3$ evolves from a classic ferroelectric to a diffuse phase transition (DPT) as x increases. Ba{[Ga$_x$,Ta$_x$]Ti$_{(1-2x)}$}O$_3$ for x ≥ 0.01 possesses diffuseness parameters comparable to Pb(Mg$_{1/3}$Nb$_{2/3}$)O$_3$-PbTiO$_3$ (PMN-PT) and recently reported (Ba$_{0.97}$Pr$_{0.03}$)(Ti$_{0.9425}$Ce$_{0.05}$)O$_3$ (BPTC), yet it lacks the frequency and temperature dependence of $T_m$ necessary to be a strictly defined relaxor ferroelectric. Additionally, Ba{[Ga$_{0.05}$,Ta$_{0.05}$]Ti$_{0.9}$}O$_3$ possesses a relative permittivity, $\varepsilon_r$, of 700±16% and dissipation factor less than 0.05 at 10 kHz within the temperature range [-75°C, 120°C]. In comparison to BaTiO$_3$, Ba{[Ga$_x$,Ta$_x$]Ti$_{(1-2x)}$}O$_3$ possesses enhanced electrical resistivity at and above room temperature. In-situ XRD, including Rietveld refinement, have been performed to determine the lattice parameter, coefficient of thermal expansion and phase transition temperature ($T_c$) of each composition within the temperature range [RT, 1000°C], thus linking the dielectric properties with the material's structure. These studies have been corroborated by temperature dependent Raman spectroscopy to compare the $T_c$ determined by electrical and structural characterization. The properties of Ba{[Ga$_x$,Ta$_x$]Ti$_{(1-2x)}$}O$_3$ are discussed in context with available models that describe donor and acceptor dopants spatially separated in the parent matrix, inter-relating lattice parameter, Curie temperature, and other material properties.

# 1. Introduction

In the recent decades, enhancing capacitor technology, including enhancing energy density, increasing the length of time such energy can be stored on capacitors, and enabling new capacitor technologies are highly desired[1]. Moreover, there is a renewed interest on lead-free barium titanate (BT) based materials after lead being reduced from many commercial materials and applications[2]. BT is the first ceramic material in which ferroelectric behavior has been observed and is presently the most widely used high permittivity ceramic material[3]. The crystal structure of BT changes in series of ferroelectric-ferroelectric phase transitions before the material enters the paraelectric region[4, 5]. The Curie temperature ($T_c \sim 131°C$ for BT as reported in the present work) is the temperature at which the tetragonal (ferroelectric) to cubic (paraelectric) phase transformation occurs on heating. BT possesses other phase transformations: orthorhombic to tetragonal (O-T) at 20°C and rhombohedral to orthorhombic (R-O) at -70°C, which are often designated $T_{c_2}$ and $T_{c_3}$, respectively. Below $T_c$, BT is distorted to a tetragonal structure with the effective polar axis along the $c$-direction. For orthorhombic and rhombohedral BT structures, the effective polar axis is along the face and body diagonal of the prototypic cubic phase, respectively.

In spite of BT having many advantages over other ferroelectric materials, it has some serious drawbacks for field tunable devices such as a narrow temperature range of operation due to variation of capacitance with temperature, low tunability (i.e. ability to controllably adjust the $\varepsilon_r$ response by varying applied electric field) except near $T_c$, and increased dissipation factor below $T_c$. Effects of dopants, substitutions and additives on the electric properties of BT have been investigated to improve the frequency adaptive performance of the material through systematic adjustment of the phase transition temperature, reduction of loss, and reduction of both temperature sensitivity and relative high permittivity while maintaining or increasing the relatively high electric-field tunability over extended temperature ranges[6–8]. Some of these property dependencies, e.g., in both dissipation factor and temperature sensitivity of the relative permittivity coincide with capacitor requirements for industry[9–12] and must coexist with high breakdown strength, relatively high permittivity, intrinsic resistor-capacitor (RC) time constant and cost of the material.

The electrical characteristics such as conductivity, $\varepsilon_r$, optical birefringence etc., of $ABO_3$ perovskite ferroelectric materials can be easily tailored by introducing specific elements in the host material in the form of substituents. A and B-site substituents in $ABO_3$ perovskite structure, typically result in a shift in $T_c$ to higher or lower temperature depending upon the substituent type and amount. Homovalent and heterovalent substitution are the cases in which one ionic species with either same or different charge is substituted, respectively, at one of the A or B perovskite sites. Dipole-like substitution refers to replacing the same parent atom (either at A- or B-site, hence, either Ba or Ti) in two adjacent or nearby unit cells with charge compensated pairs, one substituent of higher and the other substituent of lower charge, to preserve charge neutrality within the local region of the unit cells[13]. Conceptually, while mesoscopically the material remains charge neutral, a permanent dipole is created in the region, with the dipole strength depending on the charge difference and separation distance of the two substituted unit cells. Although these fields may vary in direction and magnitude, they will superpose with one another and may become quite large[14, 15]. Specifically, dipole-like substituted BT has been compared with $Ba(Ti_{0.70}Zr_{0.30})O_3$ to assess the viability of using BT based relaxors for tunable microwave applications. $BaTi_{0.90}Ga_{0.05}Nb_{0.05}O_3$ shows electric field tunability comparable to BT but has increased loss at microwave frequencies. The origin of the increased loss for such materials is unclear and it is in general considered to be dipole flipping in the form of nano clusters[7]. $Ba(Y_{0.05}, Sb_{0.05})Ti_{0.90}O_3$, for example, shows a diffuse phase transition with the $T_c$ being raised to ~200°C and does not exhibit strict relaxor behavior [16]. The main objectives of the present work are to systematically study the solubility of the Ga-Ta dipole-pair (DP) in a BT parent matrix, and investigate the effects of DP substitution and concentration on the phase evolution and dielectric properties of BT parent matrix including dielectric relaxation.

## 2. Experimental procedure

*2.1 Material synthesis*

Batches of 50 g of Ba{[$Ga_x,Ta_x$]$Ti_{(1-2x)}$}$O_3$ (*BGTx%*) with $x = 0$ (BT), 0.0025 (BGT0.25), 0.005 (BGT0.50), 0.01 (BGT1.00), 0.025 (BGT2.50) and 0.05 (BGT5.00) were prepared by ball milling the calculated and weighed stoichiometric ultra-pure $BaCO_3$, $TiO_2$, $Ga_2O_3$ and $Ta_2O_5$ oxide precursors (99.95%, 99.95%, 99.999% and 99.999% purity respectively, Alfa Aesar, Haverhill, MA, U.S.A). The precursors were mixed with isopropanol (IPA) and milled with yttria-stabilized zirconia (YSZ) grinding media in Nalgene jars using speed controlled rollers. The milling time was usually between 15 to 20 hours to achieve the desired particle size ($d_{50} \leq$ 0.5 µm) and homogeneity. The homogenous powder slurry was filtered and dried in a drying oven for 15 hours, followed by calcination (in form of pressed pellets) in air with temperature range of 1250ºC to 1300ºC and using a soaking time of 2 hours. After calcination, the calcined pellets were again ball milled in IPA with YSZ grinding media for 15 hours to eliminate particle agglomerations and to obtain fine calcined powder. The calcined powder slurries were dried and room temperature x-ray diffraction was performed for phase identification. Subsequently, the powders were loaded in a stainless-steel lined die of 30 mm inner diameter and uniaxially cold pressed at 20 MPa to form a pellet of ~ 2 mm thickness. The pressed pellets were packed and sealed in vacuum bags and iso-statically pressed at 170 MPa for 30 mins and fired. No binders or sintering aids were used to avoid additional milling and to reduce the possibility of introducing additional impurities. The sintering was performed at 1350ºC to 1500ºC depending on the composition (BT at 1300ºC and BGT5.00 at 1500ºC) for 10 hours using 5ºC /min uniform heating and cooling rates. Sintered pellets were polished to obtain parallel and smooth surfaces.

*2.2 Materials characterization*

Silver electrodes were made using DuPont 7095 low temperature air firing silver composition (DuPont chemicals, Wilmington, Delaware) on the sintered ceramic surfaces for the electrical (capacitance and resistivity) measurements. The capacitance measurements were performed over the temperature range of -195ºC to 200ºC (with 5ºC steps) and within the frequency range of 10 Hz to 1 MHz using an HP 4284 LCR meter (Hewlett-Packard Japan, Hyogo, Japan) and a temperature controlled Delta 9039 chamber (Cohu equipment, Poway, California) with multiple sample positions. Temperature dependent resistivity of materials were measured from RT to 500ºC using a Ney Qex Centurion furnace (Dentsply Sirona, York, Pennsylvania), a Keithley 6487 picoammeter and voltage source (Keithley instruments, Cleveland, Ohio) controlled using LabVIEW software with graphical user interface (GUI). The applied voltage was set at 10 to 100 V depending on the room temperature resistivity of the material composition. The structural information was obtained using a Bruker D8 Advance X-ray diffractometer system (Bruker, Karlsruhe, Germany) with Anton-Paar HTK1200N Furnace (Anton Paar, Graz, Austria) and Cu Kα X-ray radiation ($\lambda$=1.5406 Å) source. The measurements were performed over a 2-theta range of 20º to 120º and a temperature range from 30ºC to 1100ºC. Rietveld refinement was performed on the high-temperature X-ray diffraction (HT-XRD) data to determine the lattice constants, the coefficients of thermal expansion, the unit cell volume and the phase transition temperatures. Raman measurements were carried out in a LabRAM 300 spectrometer (Horiba Jobin Yvon, Villeneuve d'Ascq, France) using an excitation wavelength of $\lambda$ = 532 nm in a backscattering geometry. Temperature dependent

Raman measurements were carried out in a Linkam (THMS600) temperature controlled stage (Linkam, Tadworth, UK).

## 3. Results

*3.1 Ceramic density and macroscopic phase identification*

Cold iso-statically pressed powder compacts are ~ 60% of the theoretical density calculated by geometric density. The densities of the sintered pellets calculated by Archimedes principle are minimum of ~90% for BT and increased to a maximum of ~95% of the theoretical density for BGT5.00. Room temperature XRD of sintered samples prepared by solid-state reaction at 1500°C were performed to confirm the perovskite phase formation. No secondary phase was detected for the DP substitution level of up to x = 0.0500. BT, BGT0.25 and BGT0.50 show peak splitting in the (200) reflection and can be indexed to the tetragonal *P4mm* space group, similar to pure BT at RT. BGT1.00, BGT2.50 and BGT5.00 show broad diffraction peaks without any peak splitting, and they can be indexed, within the detection limit of the instrument, to the $Pm\bar{3}m$ space group, which is the cubic polymorph of pure BT. Thus, from room temperature XRD, a transition from tetragonal to pseudo-cubic symmetry appears to occur for compositions around and above x = 0.01. Figure 1(a) shows the room temperature XRD patterns of the investigated samples.

Figure 1 (b) provides a detailed view of the diffractogram in the vicinity of the (200) reflection. The peak near 38° consistently shifts toward lower diffraction angles (2θ) for increasing DP concentration (from 39.012° for BT to 38.884° for BGT5.00), which indicates increasing volume expansion for increasing x. With the increase in DP-(Ga,Ta) concentration, the *a* lattice parameter increases, the *c* lattice parameter decreases, the overall volume increases and the tetragonality, *c/a* ratio, decreases. At RT, the lattice parameters for pure BT are *a* = 3.9947 Å and *c* = 4.033 Å as compared to the lattice parameters of BGT0.50 of *a* = 4.001 Å and *c* = 4.026

Å. For pseudo-cubic symmetry, the lattice parameter increases with increasing DP-(Ga, Ta) concentration.

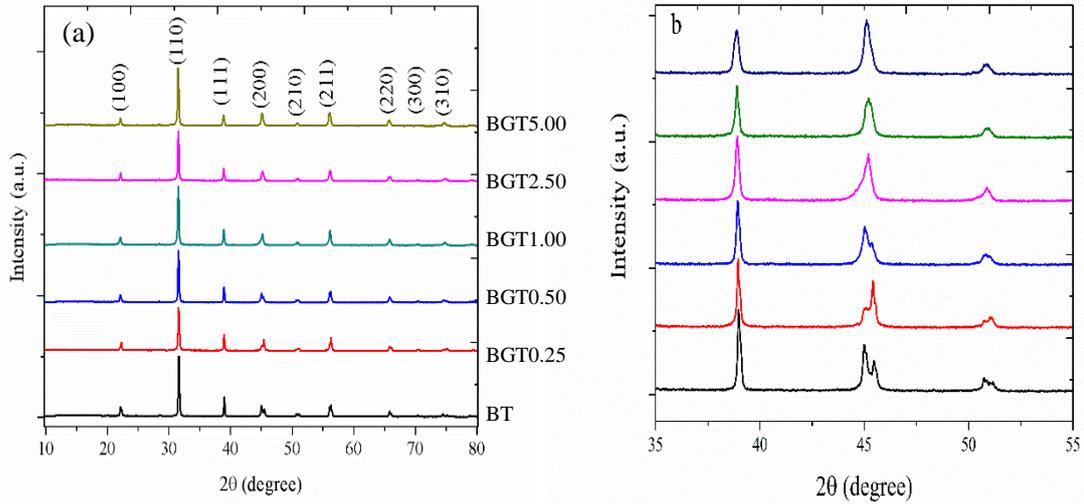

**Figure 1.** Room temperature x-ray diffraction patterns (a, b) of the samples

*3.2 Dielectric properties*

Figure 2 shows the $\varepsilon_r$ as a function of frequency [100 Hz to 1 MHz] within the temperature range [-100ºC, 195ºC] for the investigated samples. The materials present a clear evolution from a ferroelectric behavior to diffuse phase transition (DPT) behavior, finally evolving into apparent relaxor behavior for x ≥ 0.025. From figure 2, the $T_c$ as measured using $\varepsilon_r$ decreases with the increase in DP concentration from 131ºC for BT to 91ºC for BGT0.50 and eventually entering into the relaxor-like state for x ≥ 0.025 . Table 1 summarizes the observed $\varepsilon_r$ maximum and $\varepsilon_r$ at RT for BT, BGT0.25, BGT0.50, BGT1.00, BGT2.50 and BGT5.00 at 10 kHz frequency. Interestingly for $\varepsilon_r$ (cf. Figure 2), increasing the DP concentration up to 0.5 mol% does not appear to have an effect on the tetragonal-orthorhombic phase transition temperature and has only a minor effect on the cubic-tetragonal phase transition temperature, $T_c$. These results are consistent with the construct proposed by Crown et al.[14] that the charge compensated dipole pairs in the ferroelectric material will result in diffusivity of $\varepsilon_r$ while essentially not affecting the phase transition temperature(s) of the parent matrix, BT, material. The work of Crown et al. also mathematically relates the change in slope of the $\varepsilon_r$ curve in the paraelectric regime to the increase in DP concentration. However, some of the analyses are inadequate for and inconsistent in describing the entirety of experimental results presented here.

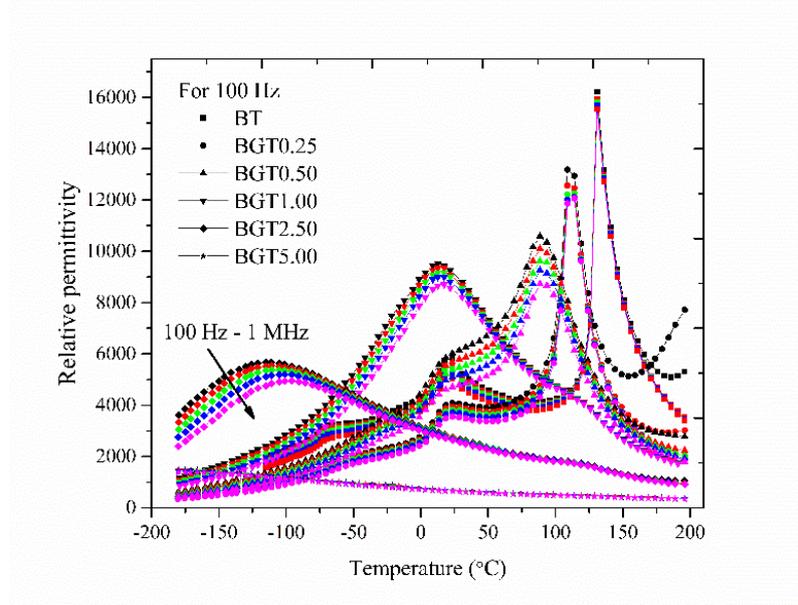

**Figure 2.** Relative permittivity as a function of frequency [100 Hz to 1 MHz] within the temperature range [-100°C, 195°C]

Table 1. The observed, $\varepsilon_r$ maximum and $\varepsilon_r$ at RT for 10 kHz frequency and $\rho$ at RT and 250°C for the investigated samples.

| Sample ID | $\varepsilon_{r\,max}$ | $\varepsilon_r$ at RT | $\rho$ at RT ($\Omega\cdot m$) | $\rho$ at 250°C ($\Omega\cdot m$) |
|---|---|---|---|---|
| BT | 16000 | 4900 | ~$10^8$ | ~$10^3$ |
| BGT0.25 | 12000 | 3800 | ~$10^9$ | ~$10^2$ |
| BGT0.50 | 9700 | 5600 | ~$10^9$ | ~$10^3$ |
| BGT1.00 | 9300 | 8600 | ~$10^{10}$ | ~$10^7$ |
| BGT2.50 | 5500 | 2600 | ~$10^{10}$ | ~$10^7$ |
| BGT5.00 | 1400 | 700 | ~$10^9$ | ~$10^7$ |

Table 1 also summarizes the resistivity (ρ) of BGTx (x= 0 to 0.05) at RT and 250°C, respectively. Initially, a drop in RT resistivity of BGT0.25 is seen when compared to BT. Once enough DP's are present, the internal electric field, which opposes the applied electric field, is strengthened and the resistivity starts to rise as seen for BGT0.50 followed by further enhancement of resistivity for BGT1.00. Continued increase of the DP concentration to 2.5 and 5 mole% results in a decrease in the RT resistivity by less than one order of magnitude and one order of magnitude respectively, which may be an indication that the local DP induced electric field is weaker than the long range electric field within the parent material. Temperature dependent resistivity measurements of the polycrystalline ceramics for the temperature range [25°C, 450°C] is also shown in Figure S1. The change in slope of resistivity for BGT1.00, BGT0.25 and BGT5.00 is an indication of the ferroelectric to paraelectric phase transition which correlates with other electrical and structural measurements and will be discussed further in the discussion section.

Table 2. The $\varepsilon_r$ observed Curie temperature ($T_c$), calculated Curie-Weiss temperature ($T_o$), Curie-Weiss Constant (C) and diffuseness parameter (γ) (calculated from Eq. 1 & 2).

| Sample ID | $T_c$ (°C) | $T_o$ (°C) | C (X $10^5$ K) | γ |
|---|---|---|---|---|
| BT | 131 | 115 | 2.18 | 1.07 |
| BGT0.25 | 112 | 100 | 1.90 | 1.06 |
| BGT0.50 | 91 | 77 | 2.41 | 1.80 |
| BGT1.00 | - | 80 | 2.01 | 1.60 |
| BGT2.50 | - | 28 | 1.46 | 1.70 |
| BGT5.00 | - | N/A | N/A | N/A |

In normal ferroelectric, above $T_c$, the temperature dependent $\varepsilon_r$ follows Curie-Weiss law. The Curie-Weiss law can be written as:

$$\frac{1}{\varepsilon} = \frac{(T-T_o)}{C} \quad (1)$$

where $T_o$ is the Curie-Weiss temperature and C is the Curie-Weiss constant.

Figure 3 (a) displays plots of $1/\varepsilon_r$ as a function of temperature at 10 kHz for BT, BGT0.25 and BGT0.50. From the figure it is evident that sample BGT0.25 follows the Curie-Weiss law just like a normal ferroelectric material, whereas BGT0.50 presents a small deviation. Figure 3(b) shows $1/\varepsilon_r$ as a function of temperature at 10 kHz frequency for BGT1.00 and BGT2.50. From the figure, it is evident that the materials BGT1.00 and BGT2.50 each possess two different linear trends. For BGT1.00, the linearity in the lower temperature response corresponds to $T_m$ at ~ 5°C in the $\varepsilon_r$ response; on the other hand, the region that follows the Curie-Weiss law at higher temperatures corresponds to the local maxima observed at ~ 125°C, which is close to the $T_c$ of BT. $T_m$ by definition is the temperature at which the maximum $\varepsilon_r$ is recorded.

In addition, the diffuseness parameter of the modified Curie-Weiss (correlation) relation is often used to distinguish diffuseness of a phase transition from a normal ferroelectric to a complete DPT, although in the presented form it does not incorporate frequency dependence. The modified Curie-Weiss relation can be written as:

$$\frac{1}{\varepsilon} - \frac{1}{\varepsilon_m} = \frac{(T-T_m)^\gamma}{C'} \qquad (2)$$

where $C'$ is the modified Curie-Weiss constant; $\gamma$ is the diffuseness parameter and $\varepsilon_m$ is relative permittivity corresponding to $T_m$. The plots of $\ln(1/\varepsilon-1/\varepsilon_m)$ versus $\ln(T-T_m)$, and the linear fit to the plots in Figure S2 provide the diffuseness parameters for all the materials and are listed in Table 2. $\gamma$ varies from 1, for a normal ferroelectric, to 2 for a complete DPT. For BGT0.50, the fit values range from 1.8 to 2.0, indicating that BGT0.50 already exhibits a complete DPT behaviour. Such DPT behaviour is comparable to recently reported $\gamma$ of 1.99 for $(Ba_{0.97}Pr_{0.03})(Ti_{0.9425}Ce_{0.05})O_3$ [17], yet there is no discernible indication of strict relaxor behavior (cf. Figure 2). Table 2 summarises the observed Curie temperature ($T_c$), calculated Curie-Weiss temperature ($T_o$), Curie Constant (C)[18] and diffuseness parameter ($\gamma$)[19] for all the samples.

The non-linear dielectric relaxation of BGT2.50 is modelled using the Vogel-Fulcher formulation,

$$f = f_o exp\left(\frac{-E_a}{k_B(T_m-T_f)}\right) \quad (3)$$

where $f_o$ is the attempt frequency, $E_a$ is the activation energy, $k_B$ is the Bolzmann constant and $T_f$ is the freezing temperature. A reasonable fit is shown in Figure S3 with parameters are $f_o$= 9.29 X $10^{13}$ Hz $E_a$=0.071 eV and $T_f$=130 K which are comparable with other BT based relaxors[20, 21]

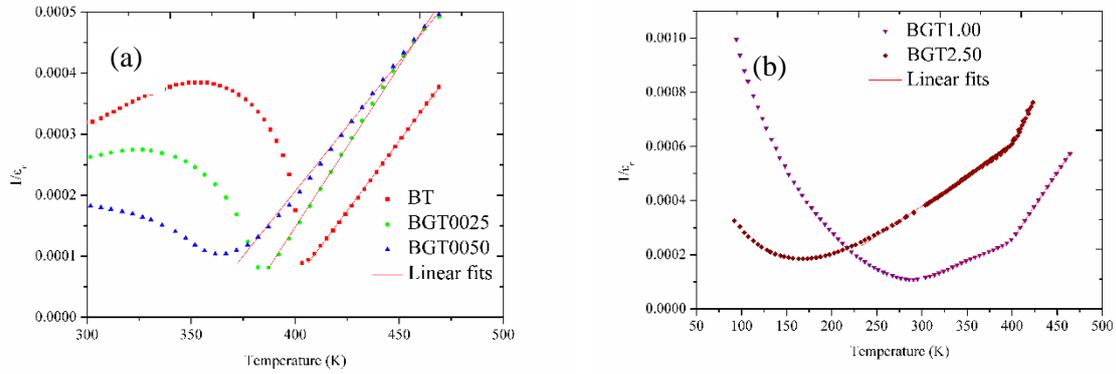

**Figure 3.** $1/\varepsilon_r$ as a function of temperature at 10 kHz for BT, BGT0.25 and BGT0.50 (a) and BGT1.00 and BGT2.50 (b)

*3.3 Temperature dependent structural characterization*

Figure 4 shows the $T_c$, $T_{c2}$ and $T_{c3}$ from $\varepsilon_r$ measurements and HTXRD for BT, BGT0.25, BGT0.50, BGT1.00 and BGT2.50. The tetragonal to cubic phase transition, otherwise known as $T_c$, are observed at 125ºC, between 100ºC to 125ºC and between 75ºC to 100ºC for BT, BGT0.25 and BGT0.50, respectively. Figure 5 shows the temperature dependent lattice parameter for BGT1.00, BGT2.50 and BGT5.00. From Figure 5, it is evident that there are two linear trend lines in the slope of each material and the connecting point of the two lines indicates a change in thermal expansion of the material. This is usually an indication of a phase transformation and it occurs for BGT1.00 and BGT2.50 at ~150°C and ~160°C (cf. Figure 4), respectively. Figures S4 and S5 show the Rietveld refinements at 75°C of BGT0.25 and BGT5.00 respectively, and are representative of the tetragonal and cubic phases as well as goodness of fit to the samples, respectively. Figure 6 shows the lattice volume expansion as a function of temperature for BT, BGT0.25, BGT0.50, BGT1.00, BGT2.50 and BGT5.00. Table 3 summarizes the crystal symmetry, the experimental unit cell dimensions (both at RT) and the calculated thermal expansion coefficient for the temperature range [200°C, 1000°C] of BT, BGT0.25, BGT0.50, BGT1.00, BGT2.50 and BGT5.00. Table S1 provides the temperature dependent lattice parameters of all the investigated samples.

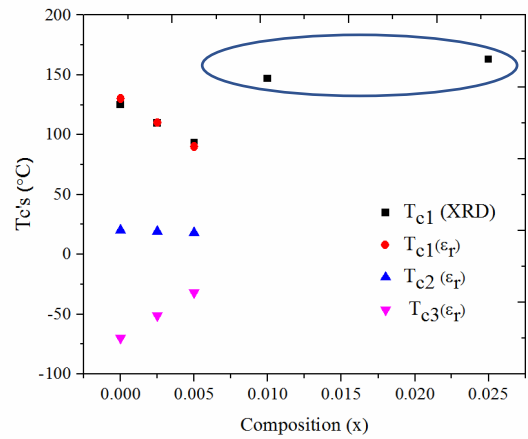

**Figure 4.** $T_{c1}$, $T_{c2}$ and $T_{c3}$ from $\varepsilon_r$ measurements and HTXRD of the samples

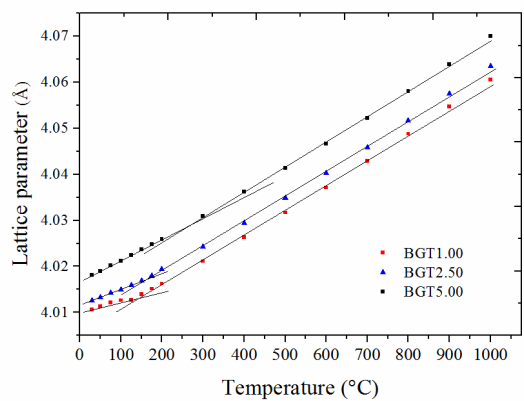

**Figure 5**. Lattice parameter as a function of temperature for BGT1.00, BGT2.50 and BGT5.00

Table 3. The crystal symmetry, the experimental unit cell dimensions (both at RT) and the calculated thermal expansion coefficient

| Sample ID | Crystal symmetry at RT | Lattice parameters at RT (Å) | | Coefficient of thermal expansion X $10^{-6}$ ($K^{-1}$) |
|---|---|---|---|---|
| | | *a* | *c* | |
| BT | Tetragonal | 3.994 | 4.033 | 13.69 |
| BGT0.25 | Tetragonal | 3.998 | 4.030 | 13.68 |
| BGT0.50 | Tetragonal | 4.001 | 4.026 | 13.71 |
| BGT1.00 | Pseudo-cubic | 4.011 | N/A | 13.57 |
| BGT2.50 | Pseudo-cubic | 4.013 | N/A | 13.52 |
| BGT5.00 | Pseudo-cubic | 4.018 | N/A | 13.41 |

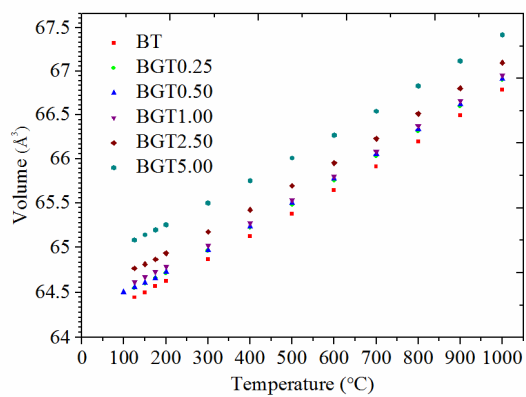

**Figure 6.** The lattice volume expansion as a function of temperature for BGTx (x=0 to 0.05)

Raman spectroscopy (RS) is a very useful technique to determine the onset of ferroelectric phases in barium titanate-based materials[22], and is able to distinguish the crossover to DPT and relaxor phases[23–25], revealing also the local displacements of Ti ions in BT systems in the cubic phase above the Curie point, where no Raman modes should be active[26]. The compositions at the crossover to relaxor state are selected for Raman investigation. Figure 7 shows the RT Raman spectra of BT, BGT0.50 and BGT1.00. All the Raman bands are broad due to the polycrystalline nature of the material (averaging of oblique phonon modes across the different crystallographic directions contained in the Raman probe[23]), and are very similar to that of BT. In particular, the interference dip at 180 cm$^{-1}$ (Mode 1), the $B_1$+E mode at 300 cm$^{-1}$ (Mode 3), the $A_1$ mode at 520 cm$^{-1}$ (Mode 4) and the high frequency LO mode at 720 cm$^{-1}$ (Mode 5) are all observed in the ferroelectric phase of BT[27]. All other modes and the nature of vibration associated with them are reported in previous papers[23]. Based on the observed Raman modes, the RT crystal symmetry of BT can be assigned to tetragonal (*T*) phase, although the spectral shape (i.e. transformation of the 180 cm$^{-1}$ dip into a peak) for BGT0.50 and BGT1.00 might indicate the presence of an orthorhombic (*O*) phase at RT in the substituted systems. It is important to note also that the asymmetric nature of mode 5 at 720cm$^{-1}$ becomes more prominent with substitution, and this is possibly due to the different $BO_6$ environment resulting from the substitution of ions of different ionic radii in the place of $Ti^{4+}$ ($Ga^{3+}$ = 0.620 Å, $Ta^{5+}$ = 0.640 Å, whereas $Ti^{4+}$ = 0.605 Å[28]). A second high frequency line at ~830 cm$^{-1}$ similar to the Nb-substituted BT is observed in BGT1.00[29]. This line appeared very sharp in Nb-substituted BT and was ascribed to a local mode induced by substitution[29]. Here the line does not appear as sharp, although the ionic radii of the substituted species are here similar to $Nb^{5+}$ (= 0.640 Å[28]). More investigations are underway in materials with higher substitution concentration in order to elucidate the origin of this mode.

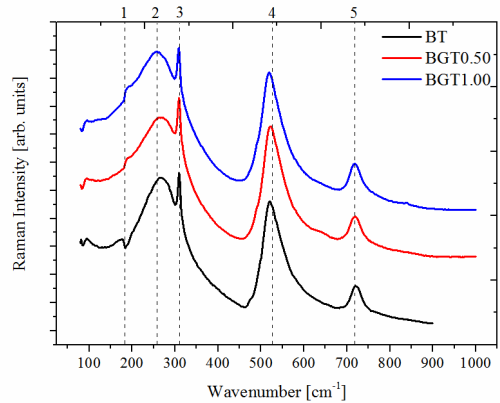

**Figure 7**. RT Raman spectrum of BT, BGT0.50 and BGT1.00 at RT.

Figure 8 shows the temperature dependent Raman spectra of BGT0.50 and BGT1.00. Tracking the spectral signature at low T and the disappearance of Raman modes with increasing temperature, the following phase evolutions have been detected: The materials possesses rhombohedral (*R*) symmetry up to ~ -50ºC , and then orthorhombic (O) until ~15ºC. Above this temperature, the materials have tetragonal (*T*) phase (disappearance of mode 4) for the whole temperature range. The phase transition sequences are in accordance with BT, and the transition temperatures match those determined by HTXRD. Interestingly, in both BGT0.50 and BGT1.00, the mode 3 at 300 cm$^{-1}$ persists well above the $T_m$ determined from dielectric measurements.

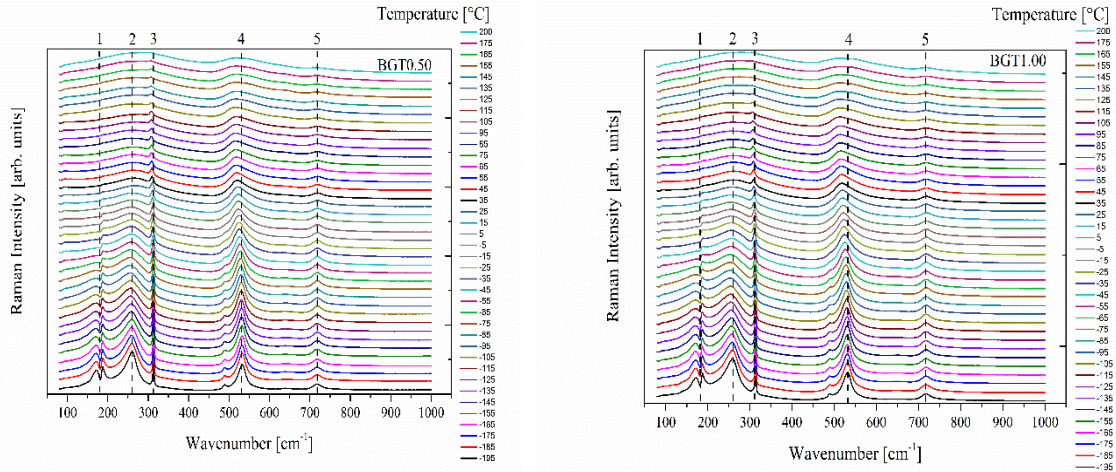

**Figure 8**. Temperature dependent Raman spectra of BGT0.50 and BGT1.00 showing ferroelectric activity well above $\varepsilon_r$ max from the permittivity measurements. Note: The spectra shown here are for every 10ºC from -195ºC to 175ºC and at 200ºC

## 4. Discussion:

As evident from Figure 1, single phase BGTx samples were fabricated up to a total substitution concentration of 10 mol%. The solubility limit of $Ga^{3+}$ is reported to be < 4 mol% in T-BT[30] and it was previously reported that the solubility limit of (Ga-Nb)-DP substituents is <30 mol%[31]. The lower solubility limit attained here could be ascribed to $Ta^{5+}$. The chemical diffusion coefficient data for heterovalent substituted BT is scarce in the literature. It is thus speculated that $Ta^{5+}$ is expected to have an even lower chemical diffusion coefficient in the BT lattice than the reported chemical diffusion coefficient of $Nb^{5+}$ [32]. The relative permittivity response (cf. Figure 3) clearly displays three different trends:

> (1) At low substitution concentration for BGT0.25, a sharp rise in the permittivity is found at the FE-PE transition, with a clear maximum at ~112ºC and a FE-like diffuseness parameter of 1.06.

(2) For BGT0.50 and BGT1.00, dramatic changes in the diffuseness parameter occur (obtained values are 1.80 and 1.60, respectively), showing a complete DPT behaviour with a clear maximum at lower temperatures: 91ºC and 25ºC, respectively.

(3) For BGT2.50, relaxor behaviour is evident, which may be the result of local polar order of different correlation length based on the chemical environment resulting from DP substitution.

In general, increasing the substitution concentration, the decreased $\varepsilon_m$ is accompanied by an increase in the diffuseness. At this point it is important to note that the room temperature permittivity of BGT1.00 is 8600 and a resistivity of $10^{10}$ Ω·m, which is significantly improved compared to BT. Such property enhancement at a very low substitution concentration can be of significant interest in energy storage applications. The resistivity of all the DP substituted samples are enhanced at RT as well as at elevated temperatures, compared to BT (cf. Table 1).

The XRD data (cf. Figures 2 and 6) indicate a non-linear volume expansion, significant compared to BT even at a very low concentration levels ($x = 0.0025$), which cannot be accounted for through changes in ionic radii, and does not obey Vegard's Law. For instance, using the room temperature six-fold coordinated Shannon ionic radii of $Ga^{3+}$ (0.620 Å) and $Ta^{5+}$ (0.640 Å) replacing $Ti^{4+}$ (0.605 Å), the lattice parameter should increase by about 0.0003, 0.0005, 0.0010, 0.0025 and 0.0050 Å for BGT0.25, BGT0.50, BGT1.00, BGT2.50 and BGT5.00, respectively, which is significantly smaller than the experimental increases of 0.0017, 0.0024, 0.004, 0.006 and 0.011 Å, respectively. This unexpected volume expansion could be explained in terms of the internal electric field induced by the DP coupling. This field results from the long-range electric field associated with ferroelectric domains superimposed on the randomly dispersed DP induced electric field. The latter would in general oppose the long-range electric field, resulting in the expansion of the effective ionic radii. Further, note that the samples BGT1.00, BGT2.50 and BGT5.00 exhibit, in addition to $T_m$, a local maximum in the $\varepsilon_r$ fairly close to the $T_c$ of pure BT of 131ºC (Such can more easily been seen by the kink in the $1/\varepsilon_r$ plot, figure 3 (b) for BGT1.00 BGT2.50 and (a) for BT). The temperature of this local maximum corresponds to the $T_c$ observed from HTXRD for these compositions (cf. Figure 4-highlighted with a circle) and roughly corresponds with the drastic change of slope of the temperature dependent resistivity (cf. figure S1). In addition, above these local maximum, the samples display Curie-Weiss behavior (cf. Figure 3b), which leads to the speculation that these local maxima are related to a long-range ferroelectric to paraelectric phase transitions in these materials. On the other hand, there is the presence of the recorded ferroelectric mode 3 at 300 cm$^{-1}$ in temperature dependent Raman spectra also well above the $T_c$ from HTXRD or the $T_m$ from dielectric measurements (cf. Figure 8). Such spectra are an indication of lattice strain – likely caused by DP defects – possibly leading to stabilization of the ferroelectric order on the nano-scale (i.e. no long-range order) even above the ferroelectric to paraelectric phase transition. However, the HTXRD data of BGT1.00, BGT2.50 and BGT5.00 were assigned to $Pm\bar{3}m$ space group. This discrepancy between the two characterization techniques can be explained in terms of the difference in the coherence length. The coherence length of Raman spectroscopy is on the order of nanometers compared to the XRD with a coherence length of ~20 unit cells in perovskite ceramics[33]. These results suggest that (i) both $T_m$ and $T_c$ coexist in these materials and (ii) the stabilization of the ferroelectric phase at high temperature is a strain-mediated mechanism through local electric-field caused by the presence of charged DP on the nanoscale. While these effects may be due to core-shell structures[7], we believe that the core-shell structures are unlikely from the material processing standpoint. In addition, the unusual volume expansion (cf. Figure 7) to the best of our knowledge have not been reported in any substituted barium titanate ceramics (substituents with similar ionic radii as Ti$^{4+}$). This volume expansion can only be explained by strong internal electric field as the result of dipole pair substituents (co-doping) and this is only valid if there is no segregation of phases. The increase in resistivity provides further evidence of the dipole pairing of substituents. It is important to note that in BT from percolation theory, core-shell is unlikely to have a significant effect in the dielectric properties until relatively high volume fraction (approximately 20% or greater) of core-shell is reached[34]. Further investigations are ongoing in order to clarify these suppositions. Note: BGT0.50 is not discussed in terms of the dielectric maximum because no $T_m$ was observed within the measured temperature range.

The dilute nanoscale dispersed DP's are expected to introduce a dipole-like interaction within the ferroelectric material matrix that is randomly distributed and is expected to have a strong interaction with the lattice compared to single-dopant homo or heterovalent substituted systems, resulting in early and rapid $\varepsilon_r$ relaxation. For instance, $Zr^{4+}$ substitution in the B-site of BT decreases the $T_c$ at a rate of ~4°C for every 1 mol% substitution concentration whereas it is ~30°C in case of $Nb^{5+}$ substitution in BT. In the current materials, DP substitution results in a decrease in $T_c$ at a rate of ~40°C (present work) for every 1 mol% substitution concentration. These values are extracted from the regions of phase diagrams that display linear relationship between the $T_c$ and substitution concentration[35, 36]. This indicates that the origin of $\varepsilon_r$ relaxation is of different mechanism in DP substituted materials compared to single-dopant homo or heterovalent substituted systems.

## 5. Conclusion:

In this work, fabrication of solid solution Ba{[$Ga_x$,$Ta_x$]$Ti_{(1-2x)}$}$O_3$ with $x$ equal to 0, 0.0025, 0.005, 0.01, 0.025 and 0.05 have been successfully demonstrated using the solid-state reaction method. Ultra-high purity barium titanate (BT) with a $T_c$ of 131°C is reported in this work. The fabricated samples were electrically and structurally characterized as function of temperature, and demonstrate typical B-site BT solid solution behavior, such as a transition from ferroelectric to diffuse phase transition (DPT) to relaxor behavior – accompanied by a decrease in the dielectric permittivity maximum for increasing substituent content. To understand the dielectric relaxation and DPT behavior Raman spectroscopy, X-ray diffraction, $\varepsilon_r$, and resistivity have been performed and discussed in detail for BGTx solid solution (x=0 to 0.05). It is important to emphasize that the relaxor behavior in a homovalent substituted system occurs at a substitution concentration of >20% in the B-site[37] and below 10% for heterovalent substituted system[36, 38], likely because of a stronger, charge-mediated mechanism giving rise to relaxor behavior. In the DP substituted system presented here, the relaxor origin is at ≤ 5% substitution concentration ($Ga^{3+}$ + $Ta^{5+}$ = 5%), which suggests that random quenched defect dipoles may have even a stronger effect in creating lattice polar disorder compared to homo or heterovalent substituted systems. It is interesting to note that the $T_m$ and $T_c$ coexist in these materials with no frequency and temperature dependent permittivity dispersion at $T_c$ for the majority of the compositions. Therefore, these materials do not fit the definition for classification as strict relaxors. More efforts in understanding the local structures and possible decoupling of DP's at high temperatures must be carried out to discover the origin of relaxor behavior and to improve the understanding of these materials.

**Acknowledgements:**

The M.S. thesis of V.K.V. would have not been possible without the support of the Kyocera Corporation, through Inamori Professorships for which S.T.M. and S.C.T. are grateful to Kyocera Corporation. V.K.V and M. D gratefully acknowledges support from the Austrian Science Fund (FWF): Project P29563-N36.

*References:*

1   X. Hao, "A review on the dielectric materials for high energy-storage application," *J. Adv. Dielectr.*, **3** [1] 1–14 (2013).

2   J. Rödel, W. Jo, K.T.P. Seifert, E.M. Anton, T. Granzow, and D. Damjanovic, "Perspective on the development of lead-free piezoceramics," *J. Am. Ceram. Soc.*, **92** [6] 1153–1177 (2009).

3   C.A. Randall and R.E. Newnham, "History of the first ferroelectric oxide, $BaTiO_3$," *Mater. Res. Inst. Pennsylvania State Univ. Univ. Park. PA 16802 USA*, (2004).

4   W.F. Forrester and R.M. Hinde, "Crystal structure of barium titanate," *Nature*, [177] 155 (1945).

5   H.F. Kay and V. P, "XCV. Symmetry changes in barium titanate at low temperatures and their relation to its ferroelectric properties," *London, Edinburgh, Dublin Philos. Mag. J. Sci.*, **40** [309] 1019–1040 (1949).

6   A.K. Tagantsev, V.O. Sherman, K.F. Astafiev, J. Venkatesh, and N. Setter, "Ferroelectric materials for microwave tunable applications," *J. Electroceramics*, **11** [1–2] 5–66 (2003).

7   A. Feteira, D.C. Sinclair, I.M. Reaney, Y. Somiya, and M.T. Lanagan, "$BaTiO_3$-Based Ceramics for Tunable Microwave Applications," *J. Am. Ceram. Soc.*, **87** [6] 1082–1087 (2004).

8   T. Maiti, R. Guo, and A.S. Bhalla, "Enhanced electric field tunable dielectric properties of $Ba Zr_x Ti_{1-x} O_3$ relaxor ferroelectrics," *Appl. Phys. Lett.*, **90** [18] 2005–2008 (2007).

9   J. Chen, "La Doping Effect on the Dielectric Property of Barium Strontium Titanate

Glass–Ceramics," *J. Mater. Sci. Technol.*, **30** [3] 295–298 (2014).

10  J. Zhi, A. Chen, Y. Zhi, P.M. Vilarinho, and L. Baptista, "Incorporation of Yttrium in Barium Titanate Ceramics," *J. Am. Ceram. Soc.*, **82** [5] 1345–1348 (1999).

11  A. Saeed, B. Ruthramurthy, W.H. Yong, O.B. Hoong, T.K. Ban, and Y.H. Kwang, "Structural and dielectric properties of iron doped barium strontium titanate for storage applications," *J. Mater. Sci. Mater. Electron.*, **26** [12] 9859–9864 (2015).

12  Z. Yao, H. Liu, Y. Liu, Z. Wu, Z. Shen, Y. Liu, and M. Cao, "Structure and dielectric behavior of Nd-doped $BaTiO_3$ perovskites," *Mater. Chem. Phys.*, **109** [2–3] 475–481 (2008).

13  V. K Veerapandiyan, W.A. Schulze, S.M. Pilgrim, S.T. Misture, and S.C. Tidrow, "Structural, electrical and spectroscopic studies of the diffuse phase transition relaxor-like ferroelectric material $Ba[(Ho,Sb)_{0.05}Ti_{0.9}]O_3$," *Ferroelectrics*, **532** [1] 168–182 (2018).

14  F.J. Crowne, S.C. Tidrow, D.M. Potrepka, and A. Tauber, "Microfields induced by random compensated charge pairs in ferroelectric materials;" pp. 131–142 in *Mater. Res. Soc.* 2002.

15  M. Deluca, Z.G. Al-Jlaihawi, K. Reichmann, A.M.T. Bell, and A. Feteira, "Remarkable impact of low $BiYbO_3$ doping levels on the local structure and phase transitions of $BaTiO_3$," *J. Mater. Chem. A*, **6** [13] 5443–5451 (2018).

16  T.R. Mion, D.M. Potrepka, F.J. Crowne, A. Tauber, and S.C. Tidrow, "Electrical and Structural Properties of $Ba(Y^{3+}, Sb^{5+})_{0.05}Ti_{0.90}O_3$," *Integr. Ferroelectr.*, **148** [1] 17–26 (2013).

17  D.Y. Lu, X.L. Gao, Y. Yuan, and A. Feteira, "High-permittivity and fine-grained ($Ba_{1-}$

$_x$Pr$_x$)(Ti$_{1-y-x/4}$Ce$_y$)O$_3$ ceramics with diffuse phase transition," *Mater. Chem. Phys.*, **228** 131–139 (2019).

18  D. Viehland, S.J. Jang, L.E. Cross, and M. Wuttig, "Deviation from Curie-Weiss behavior in relaxor ferroelectrics," *Phys. Rev. B*, **46** [13] 8003–8006 (1992).

19  S.M. Pilgrim, A.E. Sutherland, and S.R. Winzer, "Diffuseness as a Useful Parameter for Relaxor Ceramics," *J. Am. Ceram. Soc.*, **73** [10] 3122–3125 (1990).

20  L. Wu, X. Wang, and L. Li, "Lead-free BaTiO$_3$-Bi(Zn$_{2/3}$Nb$_{1/3}$)O$_3$ weakly coupled relaxor ferroelectric materials for energy storage," *RSC Adv.*, **6** [17] 14273–14282 (2016).

21  X.G. Tang, X.X. Wang, K.H. Chew, and H.L.W. Chan, "Relaxor behavior of (Ba,Sr)(Zr,Ti)O$_3$ ferroelectric ceramics," *Solid State Commun.*, **136** [2] 89–93 (2005).

22  C.H. Perry and D.B. Hall, "Temperature Dependence of the Raman Spectrum of BaTiO$_3$," *Phys. Rev. Lett.*, **15** [17] 700–702 (1965).

23  V. Buscaglia, S. Tripathi, V. Petkov, M. Dapiaggi, M. Deluca, A. Gajović, and Y. Ren, "Average and local atomic-scale structure in BaZr$_x$Ti$_{1-x}$O$_3$(x = 0.10, 0.20, 0.40) ceramics by high-energy x-ray diffraction and Raman spectroscopy," *J. Phys. Condens. Matter*, **26** [6] (2014).

24  N. Horchidan, A.C. Ianculescu, C.A. Vasilescu, M. Deluca, V. Musteata, H. Ursic, R. Frunza, B. Malic, *et al.*, "Multiscale study of ferroelectric-relaxor crossover in BaSn$_x$Ti$_{1-x}$O$_3$ ceramics," *J. Eur. Ceram. Soc.*, **34** [15] (2014).

25  M. Deluca, L. Stoleriu, L.P. Curecheriu, N. Horchidan, A.C. Ianculescu, C. Galassi, and L. Mitoseriu, "High-field dielectric properties and Raman spectroscopic investigation of the ferroelectric-to-relaxor crossover in BaSn$_x$Ti$_{1-x}$O$_3$ ceramics," *J.*

*Appl. Phys.*, **111** [8] 084102 (2012).

26   M.P. Fontana and M. Lambert, "Linear disorder and temperature dependence of Raman scattering in BaTiO$_3$," *Solid State Commun.*, **10** [1] 1–4 (1972).

27   R. Farhi, M. El Marssi, A. Simon, and J. Ravez, "A Raman and dielectric study of ferroelectric Ba(Ti$_{1-x}$Zr$_x$)O$_3$ ceramics," *Eur. Phys. J. B*, **604** 599–604 (1999).

28   R.D. Shannon, "Revised Effective Ionic Radii and Systematic Studies of Interatomic Distances in Halides and Chalcogenides," *Acta crytallographica*, **32** 751–767 (1976).

29   R. Farhi, M. El Marssi, A. Simon, and J. Ravez, "Relaxor-like and spectroscopic properties of niobium modified barium titanate," *Eur. Phys. J. B*, **18** [4] 605–610 (2000).

30   A. Feteira, D.C. Sinclair, I.M. Reaney, and S. Sheffield, "Synthesis and Characterization of BaTi$_{1-x}$Ga$_x$O$_{3-d}$ (0<x<0.15) Ceramics," *J. Am. Ceram. Soc.*, **89** [7] 2105–2113 (2006).

31   A. Feteira, D.C. Sinclair, I.M. Reaney, and M.T. Lanagan, "Structure-Property Relationships of BaTi$_{1-2y}$Ga$_y$Nb$_y$O$_3$ (0<=y<=0.35) Ceramics," *J. Am. Ceram. Soc.*, **88** [11] 3055–3062 (2005).

32   K. Kowalski, M. Ijjaali, T. Bak, B. Dupre, J. Nowotny, M. Rekas, and C.C. Sorrell, "Kinetics of Nb incorporation into barium titanate," *J. Phys. Chem. Solids*, **62** [3] 531–535 (2001).

33   R. Saito, M. Hofmann, G. Dresselhaus, and A. Jorio, "Advances in Physics Raman spectroscopy of graphene and carbon nanotubes," *Adv. Phys.*, **60** [3] 413–550 (2011).

34   P. Kim, N.M. Doss, J.P. Tillotson, P.J. Hotchkiss, M.J. Pan, S.R. Marder, J. Li, J.P. Calame, *et al.*, "High energy density nanocomposites based on surface-modified

BaTiO$_3$ and a ferroelectric polymer," *ACS Nano*, **3** [9] 2581–2592 (2009).

35    G. Canu, G. Confalonieri, M. Deluca, L. Curecheriu, M.T. Buscaglia, M. Asandulesa, N. Horchidan, M. Dapiaggi, *et al.*, "Structure-property correlations and origin of relaxor behaviour in BaCe$_x$Ti$_{1-x}$O$_3$," *Acta Mater.*, **152** 258–268 (2018).

36    R. Farhi, M.E. Marssi, A. Simon, and J. Ravez, "Relaxor-like and spectroscopic properties of niobium modied barium titanate," *Eur. Phys. J. B*, **18** 605–610 (2000).

37    G. Canu, G. Confalonieri, M. Deluca, L. Curecheriu, M.T. Buscaglia, M. Asandulesa, N. Horchidan, M. Dapiaggi, *et al.*, "Structure-property correlations and origin of relaxor behaviour in BaCe$_x$Ti$_{1-x}$O$_3$," *Acta Mater.*, **152** (2018).

38    L.P. Curecheriu, M. Deluca, Z.V. Mocanu, M.V. Pop, V. Nica, N. Horchidan, M.T. Buscaglia, V. Buscaglia, *et al.*, "Investigation of the ferroelectric-relaxor crossover in Ce-doped BaTiO$_3$ ceramics by impedance spectroscopy and Raman study," *Phase Transitions*, **86** [7] 703–714 (2013).